\documentclass[prb,preprint]{revtex4-1} 
\usepackage[utf8]{inputenc}
\usepackage[T1]{fontenc}
\usepackage{hyperref}
\hypersetup{colorlinks,linkcolor={blue},citecolor={blue},urlcolor={black}}
\usepackage{url}
\usepackage{booktabs}
\usepackage{amsfonts}
\usepackage{graphicx}
\usepackage{amsmath}
\usepackage{float}
\usepackage{mathtools}
\usepackage{tikz} 
\usetikzlibrary{arrows,scopes}
\usepackage{siunitx}
\usepackage{subcaption}
\DeclareMathOperator\erf{erf}

\newcommand{\rc}{%
\resizebox{!}{1.25ex}{%
    \begin{tikzpicture}[>=round cap]
        \clip (0.09em,-0.05ex) rectangle (0.61em,0.81ex);
        \draw [line width=.075ex, <->, rounded corners=0.12ex] (0.1em,0.1ex) .. controls (0.24em,0.4ex) .. (0.35em,0.8ex) .. controls (0.29em,0.725ex) .. (0.25em,0.6ex) .. controls (0.7em,0.8ex) and (0.08em,-0.4ex) .. (0.55em,0.25ex);
    \end{tikzpicture}%
}%
}

\newcommand{\brc}{%
\resizebox{!}{1.3ex}{%
    \begin{tikzpicture}[>=round cap]
        \clip (0.085em,-0.1ex) rectangle (0.61em,0.875ex);
        \draw [line width=.15ex, <->, rounded corners=0.12ex] (0.1em,0.1ex) .. controls (0.24em,0.4ex) .. (0.35em,0.8ex) .. controls (0.29em,0.725ex) .. (0.25em,0.6ex) .. controls (0.7em,0.8ex) and (0.08em,-0.4ex) .. (0.55em,0.25ex);
    \end{tikzpicture}%
}%
}

\newcommand{\hrc}{\hat{\brc}}

\begin{document}

\title{Elimination of Pathological Solutions of the Abraham-Lorentz Equation of Motion}

\author{Anupam Shaw}
\email{anupamshaw@outlook.com}
\affiliation{Department of Physics,\\Sardar Vallabhbhai National Institute of Technology,\\Surat, Gujarat 395007, India\\}


\begin{abstract}
For more than a century the Abraham-Lorentz equation has generally been regarded as the correct description of the dynamics of a charged particle. However, there are pathological solutions of the Abraham-Lorentz equation in which a particle accelerates in advance of the application of a force, the so-called preacceleration solutions, and solutions in which the particle spontaneously accelerates even in the absence of an external force, also known as runaway solutions. Runaways violate conservation of energy while preacceleration violates causality. In this work, I will focus on one of the most used alternative equations of motion: the Landau-Lifshitz equation, which has no pathological solution. However, it is a first-order approximation to the Abraham-Lorentz equation, raising the question of how an approximation turns out to be better than the original. Finally, some numerical results for a variety of external forces are presented to compare both the equations.
\end{abstract}

\maketitle

\section{Introduction}
Radiation has played a significant role in the development of Physics. At the fundamental level it has been crucial on deciding upon the fate of a theory. One remarkable example is that of discarding the Rutherford’s model of atom on the argument based on radiation from accelerated electrons. However, in the course of time since the establishment of electromagnetic radiation some of the most foundational issues related to it have been sidelined at the cost of immense applicability of this phenomenon. Radiation reaction is one such topic where the issues remain.

Radiation reaction is familiar to most of us through the Abraham-Lorentz equation for a radiating electron. It arises from the fact that electric and magnetic fields emitted by an accelerating electron act back on the electron, resulting in a retarding force. For more than a century the Abraham-Lorentz formula has generally been regarded as the ``correct'' non-relativistic expression for the radiation reaction. But the resulting equation of motion suffers from two problems: runaways and preacceleration.

Even if the external force is zero, the Abraham-Lorentz equation admits solutions that accelerate exponentially. These runaway solutions can be avoided by imposing suitable boundary conditions, but the cure is arguably worse than the disease, for now the particle begins to accelerate before the force is applied. Runaways violate conservation of energy, and preacceleration violates causality; both offend common sense.

Several theoretical models have been suggested to explain radiation reaction in the framework of classical electrodynamics, but the radiation reaction problem has not been completely solved. It is somewhat frustrating that
a century and a half after the first attempts, a completely satisfactory classical treatment of the reactive effects of radiation does not exist.

\section{Radiation from a Point Charge}
\subsection{What is Radiation?}
When charges accelerate, their fields can transport energy irreversibly out to infinity-a process we call radiation. Let us assume the source is localised near the origin; we would like to calculate the energy it is radiating at time $t_0$. Imagine a gigantic sphere, out at radius $r$. The power passing through its surface is the integral of the Poynting vector, $\mathbf{S}$:
\begin{equation}
    P(r,t) = \oint{\mathbf{S}\cdot d\mathbf{a}} = \frac{1}{\mu_0}\oint(\mathbf{E}\times\mathbf{B})\cdot d\mathbf{a}
\end{equation}
Because electromagnetic ``news'' travels at the speed of light, this energy actually left the source at the earlier time $t_0 = t - r/c$, so the power radiated is
\begin{equation}
    P_{rad}(t_0)=\lim_{r \to \infty}P\left(r, t_{0}+\frac{r}{c}\right)
\end{equation}
(with $t_0$ held constant). This is energy (per unit time) that is carried away and never comes back.
Now, the area of the sphere is $4\pi r^2$, so for radiation to occur the Poynting vector must decrease (at large $r$) no faster than $1/r^2$ (if it went like $1/r^3$, for example, then P would go like $1/r$, and $P_{rad}$ would be zero). According to Coulomb’s law, electrostatic fields fall off like $1/r^2$ (or even faster, if the total charge is zero), and the Biot-Savart law says that magnetostatic fields go like $1/r^2$ (or faster), which means that $S\sim 1/r^4$, for static configurations. So static sources do not radiate.
The study of radiation, then, involves picking out the parts of $\mathbf{E}$ and $\mathbf{B}$ that go like $1/r$ at large distances from the source, constructing from them the $1/r^2$ term in $\mathbf{S}$, integrating over a large spherical surface, and taking the limit as $r\rightarrow\infty$.

\subsection{Li\'enard-Wiechert Potentials}
The Li\'enard–Wiechert potentials describe the classical electromagnetic effect of a moving electric point charge in terms of a vector potential and a scalar potential in the Lorenz gauge. Built directly from Maxwell's equations, these potentials describe the complete, relativistically correct, time-varying electromagnetic field for a point charge in arbitrary motion.
The Li\'enard-Wiechert potentials $V$ (scalar potential field) and $\mathbf{A}$ (vector potential field) for a moving point charge, $q$, are given by \cite{griffiths_introduction_2018}
\begin{equation}
    V(\mathbf{r},t) = \frac{1}{4\pi \epsilon_0}\frac{q\,c}{(\rc c - \brc\cdot\mathbf{v})}
\end{equation}
and
\begin{equation}
    \mathbf{A}(\mathbf{r},t) = \frac{\mu_0}{4\pi}\frac{q\,c\,\mathbf{v}}{(\rc c - \brc\cdot\mathbf{v})} = \frac{\mathbf{v}}{c^2}V(\mathbf{r},t)
\end{equation}
where $\mathbf{v}$ is the velocity of the charge at the retarded time, and $\brc$ is the vector from the retarded position to the field point $\mathbf{r}$.

\subsection{Fields of a Moving Point Charge}
Let us calculate the electric and magnetic fields of a point
charge in arbitrary motion, using the Li\'enard-Wiechert potentials.
The electric and magnetic fields, in terms of the scalar and vector potentials are given by
\begin{equation}
    \label{eqn:EB}
    \mathbf{E} = -\mathbf{\nabla}V-\frac{\partial \mathbf{A}}{\partial t}\quad\quad\text{and}\quad\quad \mathbf{B} = \mathbf{\nabla}\times\mathbf{A}
\end{equation}
Calculating the required expressions, we get
\begin{equation}
    \mathbf{\nabla}V=\frac{1}{4\pi \epsilon_0}\frac{q\,c}{(\rc c - \brc\cdot\mathbf{v}\,)^3}\left[(\rc c - \brc\cdot\mathbf{v}\,)\mathbf{v}-(c^2-v^2+\brc\cdot\mathbf{a}\,)\brc\right]
\end{equation}
\begin{equation}
    \frac{\partial\mathbf{A}}{\partial t}=\frac{1}{4\pi \epsilon_0}\frac{q\,c}{(\rc c - \brc\cdot\mathbf{v}\,)^3}\left[(\rc c-\brc\cdot\mathbf{v}) \left(-\mathbf{v}+\frac{\rc}{c}\mathbf{a}\right)+\frac{\rc}{c}(c^2-v^2+\brc\cdot\mathbf{a})\mathbf{v}\right]
\end{equation}
and
\begin{equation}
    \mathbf{\nabla}\times\mathbf{A}=-\frac{q}{4\pi\epsilon_0 c}\frac{1}{(\mathbf{u}\cdot\brc)^3}\brc\times\left[(c^2-v^2)\mathbf{v}+(\brc\cdot\mathbf{a})\mathbf{v}+(\brc\cdot\mathbf{u})\mathbf{a}\right]
\end{equation}
Here $\mathbf{a}\equiv\mathbf{\dot{v}}$ is the acceleration of the particle at retarded time and $\mathbf{u}\equiv c\,\hrc-\mathbf{v}$.
After substituting these expressions in Equation (\ref{eqn:EB}), we get
\begin{equation}\label{eqn:E}
    \mathbf{E}(\mathbf{r},t)=\frac{q}{4\pi\epsilon_0}\frac{\brc}{(\brc\cdot\mathbf{u}\,)^3}\left[\left(c^2-v^2\right)\mathbf{u}+\brc\times\left(\mathbf{u}\times\mathbf{a}\,\right)\right]
\end{equation}
and
\begin{equation}\label{eqn:B}
    \mathbf{B}(\mathbf{r},t)=\frac{1}{c}\hrc\times\mathbf{E}(\mathbf{r},t)
\end{equation}
The first term in $\mathbf{E}$ falls off as the inverse square
of the distance from the particle. Because it does not depend on the acceleration, it is also known as the \emph{velocity field}. The second term falls off as the inverse first power of $r$ and is therefore dominant at large distances. As we shall soon see, it is this term that is responsible for electromagnetic radiation; accordingly, it is called the \emph{radiation field}, or, since it is proportional to $a$, the \emph{acceleration field}.

\subsection{Power Radiated by a Point Charge}
The fields of a point charge $q$ in arbitrary motion are given by Equations (\ref{eqn:E}) and (\ref{eqn:B}). The Poynting vector in this case becomes
\begin{equation}\label{eqn:poynting}
    \begin{split}
        \mathbf{S}=\frac{1}{\mu_0}(\mathbf{E}\times\mathbf{B})&=\frac{1}{\mu_0 c}[\mathbf{E}\times(\hrc\times\mathbf{E})]\\
        & =\frac{1}{\mu_0 c}[E^2\hrc-(\hrc\cdot\mathbf{E})\mathbf{E}]
    \end{split}
\end{equation}
However, not all of this energy flux constitutes radiation; some of it is just field energy carried along by the particle as it moves. The radiated energy is the stuff that, in effect, detaches itself from the charge and propagates off to infinity.
\begin{figure}[H]
    \centering
    \includegraphics[width=0.3\textwidth]{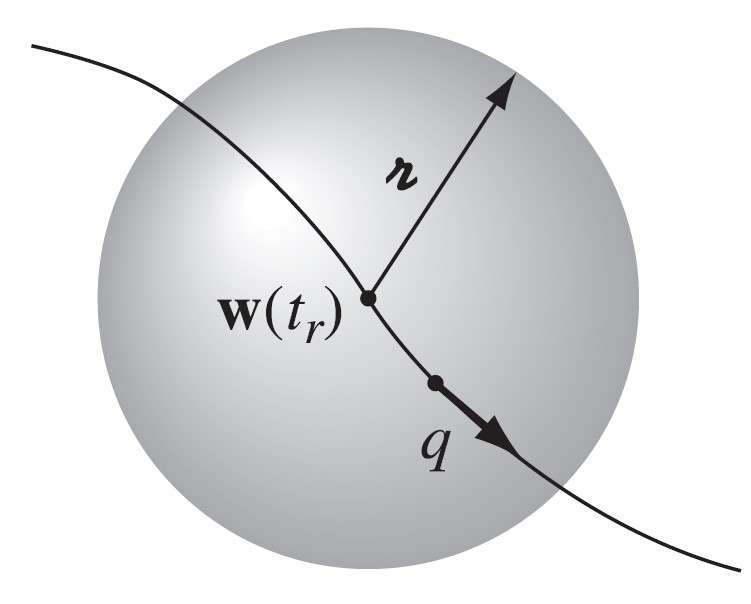}
    \label{fig:sphere}
    \caption{A sphere centered at the position of the particle at time $t_r$.}
\end{figure}

To calculate the total power radiated by the particle at time $t_r$, let us draw a huge sphere of radius $\rc$, centered at the position of the particle (at time $t_r$), wait the appropriate interval
\begin{equation}
    t-t_r=\frac{\rc}{c}
\end{equation}
for the radiation to reach the sphere, and at that moment integrate the Poynting vector over the surface.
Now, the area of the sphere is proportional to $\rc^2$, so any term in $\mathbf{S}$ that goes like $\rc^{-2}$ will yield a finite answer, but terms like $\rc^{-3}$ or $\rc^{-4}$ will contribute nothing in the limit $\rc\rightarrow\infty$. For this reason, only the acceleration fields represent true radiation (hence their other name, radiation fields):
\begin{equation}\label{eqn:rad_field}
    \mathbf{E}=\frac{q}{4\pi\epsilon_0}\frac{\rc}{(\brc\cdot\mathbf{u})^3}[\brc\times(\mathbf{u\times\mathbf{a}})]
\end{equation}
The velocity fields carry energy and as the charge moves this energy is dragged along but it’s not radiation. Now $\mathbf{E}_{rad}$ is perpendicular to $\hrc$, so the second term in Equation (\ref{eqn:poynting}) vanishes and we get
\begin{equation}
    \mathbf{S}_{rad}=\frac{1}{\mu_0 c}E_{rad}^2\,\hrc
\end{equation}

\subsection{Larmor Formula}
If the charge is instantaneously at rest (at time $t_r$), then $\mathbf{u}=c\hrc$, and Equation (\ref{eqn:rad_field}) becomes
\begin{equation}
    \mathbf{E}_{rad}=\frac{q}{4\pi\epsilon_0 c^2\rc}[\hrc\times(\hrc\times\mathbf{a})]=\frac{\mu_0 q}{4\pi\rc}[(\hrc\cdot\mathbf{a})\hrc-\mathbf{a}]
\end{equation}
In that case
\begin{equation}
    \mathbf{S}_{rad}=\frac{1}{\mu_{0}c}\left(\frac{\mu_{0}q}{4\pi\rc}\right)^2\left[a^2-\left(\hrc\cdot\mathbf{a}\,\right)^2\right]\hrc
\end{equation}
If the angle between $\hrc$ and $\mathbf{a}$ is assumed to be $\theta$, we have
\begin{equation}
    \mathbf{S}_{rad}=\frac{\mu_0 q^2 a^2}{16\pi^{2}c}{\left(\frac{\sin^2{\theta}}{\rc^2}\right)}\hrc
\end{equation}
The total power radiated by the accelerated charge at time $t$ is
\begin{equation}
    P=\oint\mathbf{S}_{rad}\cdot d\mathbf{a}={\frac{\mu_0 q^2 a^2}{16\pi^2 c}}\int{\frac{sin^2{\theta}}{\rc^2}}\rc^2 \sin{\theta}\,d\theta\,d\phi
\end{equation}
After integrating, we get the famous Larmor power formula:
\begin{equation}\label{eqn:larmor}
    P=\frac{\mu_0 q^2}{6\pi c} a^2
\end{equation}
In the context of special relativity, the condition $v=0$ simply represents an astute choice of reference system, with no essential loss of generality. If we can decide how $P$ transforms, we can deduce the general (Li\'enard) result from the $v=0$ (Larmor) formula \cite{bhattacharya_introduction_2021}.

\section{Radiation Reaction}
A particle can accelerate only when some external force $\mathbf{F}_{ext}$ acts on it for some finite time. In reality as the external force accelerates the charged particle, the particle starts to radiate electromagnetic radiation. As electromagnetic radiation carries away energy from the charged particle, it is natural to think that the charged particle loses energy while accelerating. As the charged particle loses energy due to radiation, the velocity of the charged particle will decrease. The decelerating effect on the charge can be assumed to be produced by some force acting on the charge $\mathbf{F}_{rad}$. This force $\mathbf{F}_{rad}$ is called the force due to radiation reaction.
\subsection{Abraham-Lorentz Equation of Motion}
The simplest way to calculate the radiation reaction is by exploiting conservation of energy. The work done by the radiation reaction force $\mathbf{F}_{rad}$ over the time interval $t_1<t<t_2$ and the energy radiated by the particle over that time interval must balance and it follows
\begin{equation}\label{eqn:radreac}
    \int_{t_1}^{t_2}\mathbf{F}_{rad}\cdot\mathbf{v}\,dt = -\int_{t_1}^{t_2}P(t)\,dt
\end{equation}
Here $P(t)$ is the Larmor power expression:
\begin{equation}
    P=\frac{\mu_0 q^2}{6\pi c} a^2 = m\tau|\mathbf{a}|^2,
\end{equation}
where $m$ is the mass of the particle and
\begin{equation}
    \tau=\frac{\mu_0q^2}{6\pi m c}.
\end{equation}
Rewriting the right-hand side of Equation (\ref{eqn:radreac}):
\begin{equation}
    \int_{t_1}^{t_2}\mathbf{F}_{rad}\cdot\mathbf{v}\,dt = -m\tau\int_{t_1}^{t_2}(\mathbf{a}\cdot\mathbf{a})\,dt
\end{equation}
and integrating by parts leads to
\begin{equation}\label{eqn:intparts}
    \int_{t_1}^{t_2}\mathbf{F}_{rad}\cdot\mathbf{v}\,dt = -m\tau\mathbf{a}\cdot\mathbf{v}\Big|_{t_1}^{t_2}+m\tau\int_{t_1}^{t_2}\mathbf{\dot{a}}\cdot\mathbf{v}\,dt
\end{equation}
The boundary term vanishes if the motion is periodic or if the situation is such that the velocity of the particle becomes perpendicular to the acceleration vector at $t_1$ or $t_2$. Then we can write
\begin{equation}
    \int_{t_1}^{t_2}\left(\mathbf{F}_{rad}-m\tau\mathbf{\dot{a}}\right)\cdot\mathbf{v}\,dt=0
\end{equation}
For the integral to vanish for all pertinent cases in general the integrand has to vanish. As in the general case the velocity of the particle in motion is non-zero we must have
\begin{equation}\label{eqn:abraham}
    \mathbf{F}_{rad}=m\tau\mathbf{\dot{a}}
\end{equation}
This radiation reaction force is called the \emph{Abraham-Lorentz force}. It can be seen that the radiation reaction force vanishes if $\mathbf{\dot{v}}$ vanishes for an extended time period as in that case $\mathbf{\ddot{v}}$ also vanishes. The force is also proportional to $q^2$. We can now write the equation of motion for the radiating charged particle
\begin{equation}\label{eqn:eom}
    m\mathbf{a}=\mathbf{F}_{ext}+\mathbf{F}_{rad}
\end{equation}
as
\begin{equation}\label{eqn:al-eom}
    m\mathbf{a}=\mathbf{F}_{ext}+m\tau\mathbf{\dot{a}}
\end{equation}
or
\begin{equation}
    m(\mathbf{a}-\tau\mathbf{\dot{a}})=\mathbf{F}_{ext}
\end{equation}
A surprising feature of the Abraham-Lorentz equation is that it involves the derivative of the acceleration vector. The equations of motion are therefore third-order differential equations for $\mathbf{x}(t)$, a very unusual situation that necessitates some reflection. In the usual case of second-order differential equations, the initial data consists of the particle’s position and velocity at $t=0$, and this information is sufficient to provide a unique solution. A third-order equation requires more, however, and it is not clear \emph{a priori} what the additional piece of initial data should be. To see the peculiar nature of the Abraham-Lorentz equation we see that this equation predicts radiation even when there is no external force $\mathbf{F}_{ext}$.
\subsection{Runaway Solutions}
Suppose a particle is subject to no external forces, then the Abraham-Lorentz equation gives
\begin{equation}
    \mathbf{a}=\tau\mathbf{\dot{a}}
\end{equation}
If this equation had only one solution, $\mathbf{a=0}$ and $\mathbf{\dot{a}=0}$ then everything should have been fine. The above solution says that the charged particle is at rest or moves with uniform velocity such that there is no electromagnetic radiation. In absence of any external force, this is a fine solution. The difficulty is that this equation does have another nontrivial solution as
\begin{equation}
    \mathbf{a}=\mathbf{a}_0\,e^{t/\tau}
\end{equation}
where $\mathbf{a}_0$ is the acceleration of the particle at $t=0$. This solution grows indefinitely with time. This kind of a solution is generally called a runaway solution which wildly increases without bound. Moreover, this solution is problematic in another sense, this solution does not satisfy the property $\mathbf{v\cdot a=0}$ at any time.
Consideration of the time constant $\tau$ for an electron $(\tau\sim10^{-23})$ immediately shows that runaway solutions
are totally unphysical; an electron at rest would respond to an external perturbation by accelerating to extraordinarily high speeds over very short time scales. Such behaviour is generic, and not a peculiarity of vanishing external fields: a runaway term can always be added to any solution of Equation (\ref{eqn:al-eom}), and its rapid growth implies it will dominate any acceleration generated by the applied fields. Runaway solutions must be excluded on physical grounds.
\subsection{Rohrlich's Integro-Differential Equation}
Suppose that the external force only depends on time so that we can write the basic equation of motion as
\begin{equation}\label{eqn:al-eom1}
    m(\mathbf{a}-\tau\mathbf{\dot{a}})=\mathbf{F}_{ext}(t)
\end{equation}
To tackle the problem of the runaway solution of the Abraham-Lorentz equation, one can transform the above equation into an integro-differential equation \cite{rohrlich_classical_2007}. We can write the general solution to Equation (\ref{eqn:al-eom1}) as
\begin{equation}\label{eqn:sol1}
    \mathbf{a}\,(t)=e^{t/\tau}\left[\mathbf{a}(t_0)\,e^{-t_0/\tau}-\frac{1}{m\tau} \int_{t_0}^{t}\mathbf{F}_{ext}(t')\,e^{-t'/\tau}\,dt'\right],
\end{equation}
where it is assumed the acceleration of the charged particle to be momentarily zero at $t_0$, and $\mathbf{F}_{ext}(t)$ goes to zero in the infinite past, sufficiently rapidly that the integral is well defined. The constant vector $\mathbf{a}(t_0)$ is not constrained \emph{a priori}; this is the third piece of data that must be specified to completely determine the motion of the particle. 
Runaway behaviour can be eliminated by demanding that the acceleration tends to zero in the asymptotic future, once all forces have finished acting. This can be implemented in Equation (\ref{eqn:sol1}) by taking the limits $t_0\rightarrow\infty$, $\mathbf{a}(t_0)\rightarrow0$ so that
\begin{equation}
    \mathbf{a}\,(t)=\frac{1}{m\tau} \int_{t}^{\infty}\mathbf{F}_{ext}(t')\,e^{-(t'-t)/\tau}\,dt'
\end{equation}
Applying the change of variable $s=(t'-t)/\tau$ yields the Rohrlich's integro-differential equation \cite{rohrlich_electron:_1973}:
\begin{equation}\label{eqn:intdiff}
    \mathbf{a}\,(t)=\frac{1}{m} \int_{0}^{\infty}\mathbf{F}_{ext}(t+\tau s)\,e^{-s}\,ds
\end{equation}
\subsection{Acausal Preacceleration}
It is evident that Equation (\ref{eqn:intdiff}) exhibits an unphysical phenomenon called preacceleration. The acceleration at time $t$ depends on the applied force at all subsequent times. The removal of runaway solutions thus requires that the particle is prescient. This fact makes the law of motion acausal. Although preacceleration may be eliminated and causality restored if $t_0\rightarrow-\infty$ is chosen in Equation (\ref{eqn:sol1}) instead of $t_0\rightarrow\infty$, runaway behaviour re-emerges.
Consider a particularly simple case, in which the external force is turned on abruptly at $t=0$ and stays constant thereafter:
\begin{equation}
    \mathbf{F}_{ext}(t)=\mathbf{F}_0\,\theta(t),
\end{equation}
where $\mathbf{F}_0$ is a constant vector and $\theta(t)$ is the Heaviside step function.
The integro-differential equation of motion then gives
\begin{equation}
    \mathbf{a}(t)=\frac{\mathbf{F}_0}{m}\int_{0}^{\infty}\theta(t+\tau s)\,e^{-s}\,ds
\end{equation}
After evaluating the integral \cite{poisson_introduction_1999}, we get
\begin{equation}
    \mathbf{a}(t)=\frac{\mathbf{F}_0}{m}\left[\,\theta(t)+\theta(-t)\,e^{t/\tau}\right]
\end{equation}
Although $\mathbf{a}$ is now sensibly behaved for $t>0$, we see that its behaviour is rather strange for $t<0$: At a time $\sim\tau$ prior to the time at which the external force switches on, the acceleration begins to increase. This is the problem of preacceleration. It seems like the particle knows what force it will experience in the future and adjusts its acceleration accordingly. This is a clear violation of the principle of causality. In other words, \emph{the effect cannot precede the cause.}
\begin{figure}[H]
\centering
\includegraphics[width=0.5\textwidth]{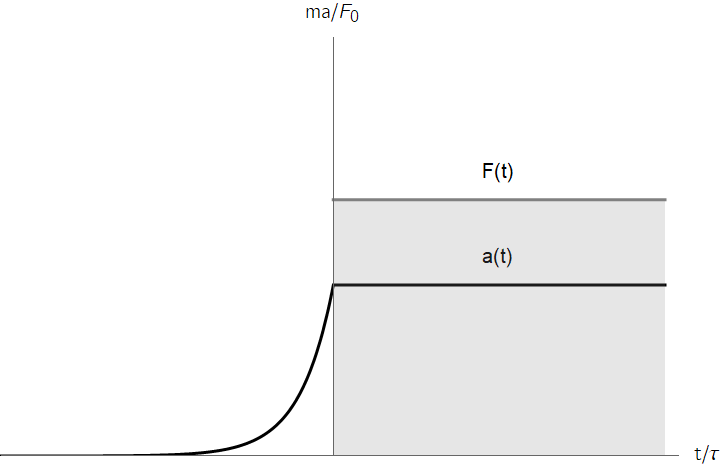}
\label{fig:preaccn}
\caption{Acceleration of a charged particle (dark line) under the action of an external force (shaded region) that is turned on abruptly at $t=0$ and remains constant thereafter. The charge, originally at rest, gets a motion (preacceleration) before the force acts.}
\end{figure}
The preacceleration phenomenon is not an artifact of the Heaviside step function, not a consequence of the fact that we are working at the moment in the non-relativistic approximation\dots but a systemic feature of the classical self-interaction problem. Roughly, preacceleration may be considered to arise because the leading edge of the extended classical source makes advance contact with the force field.
The characteristic preacceleration time is small, being given by $\tau$ ($\sim10^{-23}$ seconds for an electron). Preacceleration represents a microscopic violation of causality\dots and so it is, but the phenomenon lies so deep within the quantum regime as to be (or so I believe) classical unobservable in every instance. Preacceleration is generally considered to be (not a physical but) a merely ``mathematical phenomenon,'' a symptom of an attempt to extend classical physics beyond its natural domain of applicability.
As Wheeler and Feynman wrote, ``Preacceleration and the force of radiation reaction which calls it forth are both departures from that point of view of nature for which one once hoped, in which the movement of a particle at a given instant would be completely determined by the motions of all other particles at earlier moments.''\cite{wheeler_interaction_1945}

\section{An Alternative Equation of Motion}
Given the difficulties facing the Abraham-Lorentz equation, a number of researchers have proposed alternative theories \cite{burton_aspects_2014,yaghjian_relativistic_2006} to describe the response of a particle to its emission of radiation. Although none of these has achieved widespread acceptance, it is of interest to explore the various motivations that led to a number of them, along with their respective advantages and pitfalls. In this work, I will specifically discuss about one of the most accepted description of the dynamics of a charged particle when the external field is sufficiently weak and slowly varying in space and time.
\subsection{Landau-Lifshitz Equation of Motion}
Runaways and preacceleration can be removed simultaneously by reducing the order of the Abraham-Lorentz equation \cite{griffiths_abrahamlorentz_2010}. In the first instance we ignore the radiation reaction altogether since it typically amounts to a small perturbation. So Equation (\ref{eqn:eom}) says
\begin{equation}
    \mathbf{F}_{ext}=m\mathbf{a},
\end{equation}
and hence
\begin{equation}
    \mathbf{\dot{a}}=\frac{1}{m}\mathbf{\dot{F}}_{ext}.
\end{equation}
Then substitute this in Equation (\ref{eqn:abraham}) to calculate the radiation reaction:
\begin{equation}\label{eqn:ll_rr}
    \mathbf{F}_{rad}=\tau\mathbf{\dot{F}}_{ext}.
\end{equation}
Landau and Lifshitz introduced this expression as an approximation to Abraham-Lorentz formula \cite{landau_classical_1971,jackson_classical_1999}. One can think of the Landau–Lifshitz formula as the first term in a perturbative expansion of the Abraham–Lorentz formula. Now, the equation of motion for a radiating point charge becomes
\begin{equation}\label{eqn:ll-eom}
    m\mathbf{a}=\mathbf{F}_{ext}+\tau\mathbf{\dot{F}}_{ext}.
\end{equation}
This equation has no third derivative of position and, unlike the Abraham-Lorentz equation, does not suffer from the same pathologies as the Abraham-Lorentz equation. Therefore, the Landau-Lifshitz equation is generally accepted to be the correct classical equation of motion for a non-relativistic charged point particle if the external fields are sufficiently weak and vary sufficiently slowly (in position and time).
To get an idea of the range of validity of the Landau-Lifshitz equation, consider a sinusoidal force $f(t)=f_{0}\exp(-i\omega t)$, in which case
\begin{equation}
    \Bigg|\frac{\tau\dot{f}(t)}{f(t)}\Bigg|=\omega\tau\approx\frac{\omega r_0}{c}\approx\frac{1}{137}\frac{\hbar\omega}{mc^2}
\end{equation}
If this is not very small, we must be dealing with frequencies $\omega$ for which the classical theory breaks down. If it is very small, Equation (\ref{eqn:ll-eom}) is at least compatible with classical non-relativistic theory. It has been shown \cite{hammond_radiation_2008} that this condition fails at optical frequencies as the intensity climbs above $10^{23}\,\si{W cm^{-2}}$.
Ideally, we would like a formula for the radiation reaction that holds for any external force. Attempts to find an alternative classical equation of motion for the point charge, one that is free from pathologies and fully consistent with energy-momentum conservation and Maxwell’s equations, so far appear unsuccessful. Nevertheless, the approximation of the Landau-Lifshitz equation can give useful solutions as long as the precautions discussed above are heeded \cite{baylis_energy_2002}.

\section{Numerical Results}
In this section I will consider different forms of the external force acting on an accelerating charged particle and observe the resulting acceleration with respect to time based on the two models of the radiation reaction phenomenon that I have already discussed: the Abraham-Lorentz equation of motion (\ref{eqn:al-eom}) and the Landau-Lifshitz equation of motion (\ref{eqn:ll-eom}). I will compare the two models to determine which one is superior.

Consider a rectangular form of an external force such that
\begin{equation}\label{eq:rect}
    F_{ext}(t)= \begin{cases}
    0, & t\leq 0 \\
    F, & 0<t<T \\
    0, & t\geq T.
    \end{cases}
\end{equation}
The Abraham-Lorentz equation gives
\begin{equation}
    a(t)=\begin{dcases}
    \dfrac{F}{m}(1-e^{-T/\tau})e^{t/\tau}, & t\leq 0 \\
    \dfrac{F}{m}(1-e^{(t-T)/\tau}), & 0<t<T \\
    0, & t\geq T.
    \end{dcases}
\end{equation}
The Landau-Lifshitz equation gives
\begin{equation}
    a(t)=\begin{dcases}
    0, & t\leq 0 \\
    \dfrac{F}{m}, & 0<t<T \\
    0, & t\geq T,
    \end{dcases}
\end{equation}
together with delta functions at $0$ and $T$,
\[\tau\frac{F}{m}[\delta(t)-\delta(t-T)].\]
\begin{figure}[H]
\centering
    \begin{subfigure}[b]{0.5\textwidth}            
            \includegraphics[width=\textwidth]{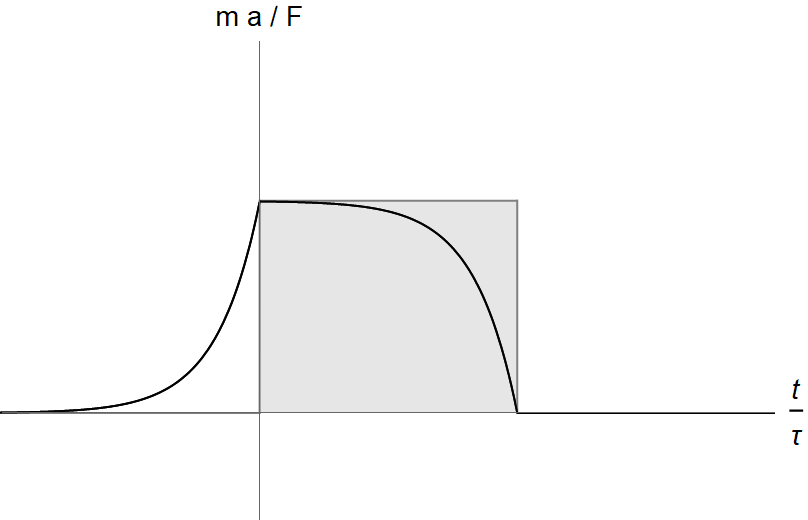}
            \caption{Abraham-Lorentz}
            \label{fig:AL_rect}
    \end{subfigure}%
    \begin{subfigure}[b]{0.5\textwidth}
            \centering
            \includegraphics[width=\textwidth]{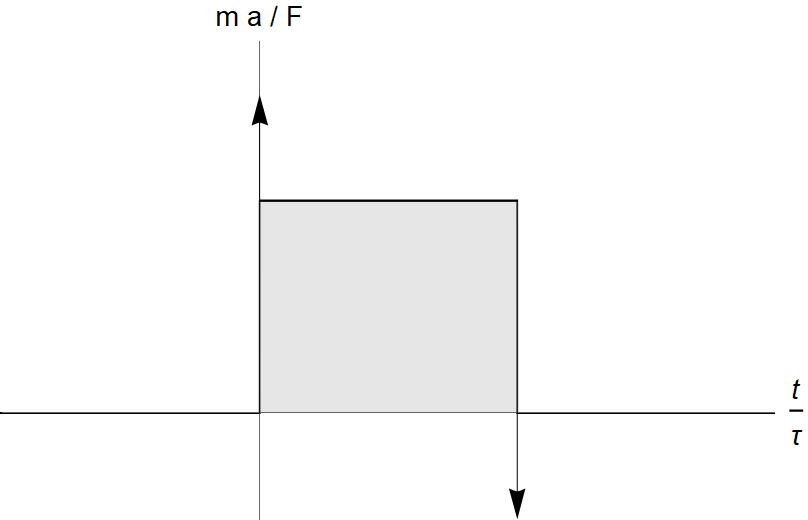}
            \caption{Landau-Lifshitz}
            \label{LL_rect}
    \end{subfigure}
    \caption{Acceleration under a rectangular force. The arrows in the Landau-Lifshitz plot indicate the two delta functions, one at $t=0$ and one at $t=T$. The shaded areas indicate the external force.}\label{fig:rect}
\end{figure}
Now, consider a Gaussian force:
\begin{equation}
    F_{ext}(t)=F e^{-t^2/T^2}.
\end{equation}
The Abraham-Lorentz equation gives
\begin{subequations}
\begin{align}
    a(t)&=\frac{F}{m\tau}e^{t/\tau}\int_{t}^{\infty}e^{-t'/\tau}e^{-{t'}^2/T^2}\,dt'\\[3mm]
    &=\frac{\sqrt{\pi}F T}{2m\tau}e^{T^2/4\tau^2}\left[1-\erf\left(\frac{t}{T}+\frac{T}{2\tau}\right)\right]e^{t/\tau},
\end{align}
\end{subequations}
while the Landau-Lifshitz equation gives
\begin{equation}
    a(t)=\frac{F}{m}\left(1-\frac{2\tau t}{T^2}\right)e^{-t^2/T^2}.
\end{equation}
\begin{figure}[H]
\centering
    \begin{subfigure}[b]{0.5\textwidth}            
            \includegraphics[width=\textwidth]{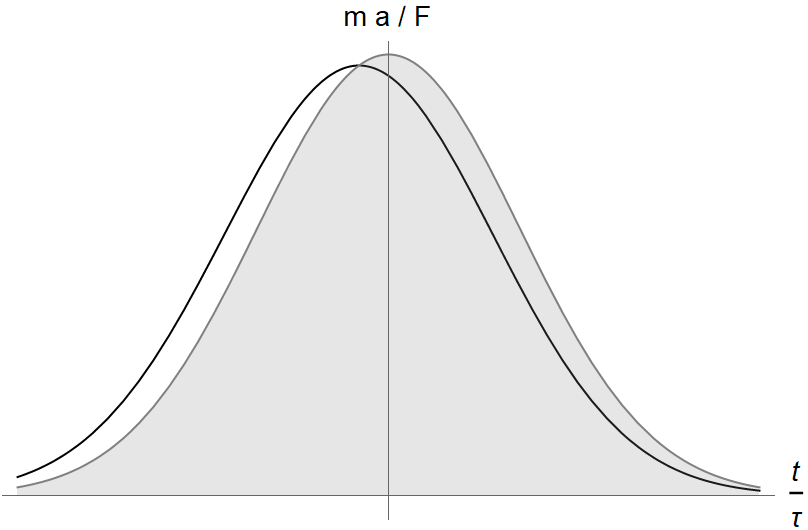}
            \caption{Abraham-Lorentz}
            \label{fig:AL_gauss}
    \end{subfigure}%
    \begin{subfigure}[b]{0.5\textwidth}
            \centering
            \includegraphics[width=\textwidth]{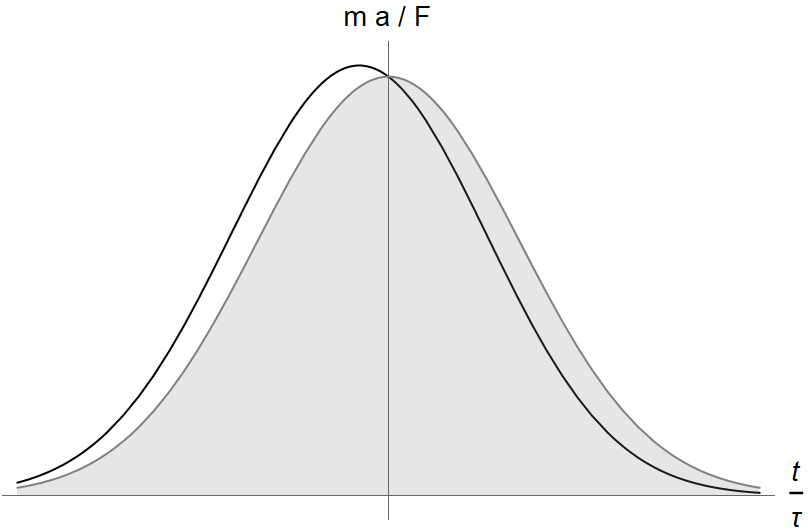}
            \caption{Landau-Lifshitz}
            \label{LL_gauss}
    \end{subfigure}
    \caption{Acceleration under a Gaussian force.}\label{fig:gauss}
\end{figure}
Finally, consider an exponential force:
\begin{equation}
    F_{ext}(t)=F e^{-|t|/T}.
\end{equation}
The Abraham-Lorentz equation gives
\begin{subequations}
\begin{align}
    a(t)&=\frac{F}{m\tau}e^{t/\tau}\int_{t}^{\infty}e^{-t'/\tau}e^{-|t'|/T}\,dt'\\[3mm]
    &=\begin{dcases}
    \dfrac{F}{m(T-\tau)}\left(e^{t/T}-\frac{2\tau}{T+\tau}e^{t/\tau}\right), & t<0 \\
    \dfrac{F}{m(T+\tau)}e^{-t/T}, & t>0.
    \end{dcases}
\end{align}
\end{subequations}
The Landau-Lifshitz equation gives
\begin{align}
    a(t)=\begin{dcases}
    \dfrac{F}{m}\left(1+\frac{\tau}{T}\right)e^{t/T}, & t<0 \\
    \dfrac{F}{m}\left(1-\frac{\tau}{T}\right)e^{-t/T}, & t>0.
    \end{dcases}
\end{align}
\begin{figure}[H]
\centering
    \begin{subfigure}[b]{0.5\textwidth}            
            \includegraphics[width=\textwidth]{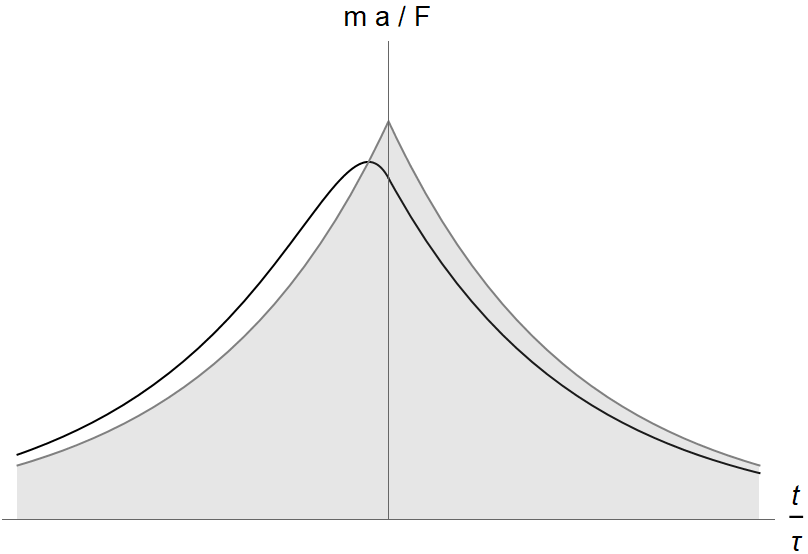}
            \caption{Abraham-Lorentz}
            \label{fig:AL_exp}
    \end{subfigure}%
    \begin{subfigure}[b]{0.5\textwidth}
            \centering
            \includegraphics[width=\textwidth]{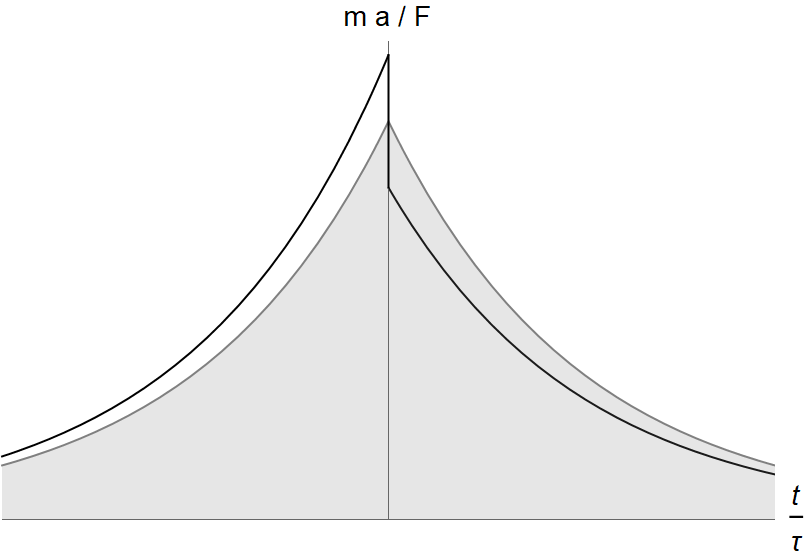}
            \caption{Landau-Lifshitz}
            \label{LL_exp}
    \end{subfigure}
    \caption{Acceleration under an exponential force.}\label{fig:exp}
\end{figure}

\section{Conclusions}
For each external force (rectangular, Gaussian, and exponential), I have solved the corresponding Abraham-Lorentz equation and the Landau-Lifshitz equation to get the acceleration of the radiating charged particle. Then I have plotted the acceleration of the particle with respect to time and tried to ascertain which equation of motion correctly describes the dynamics of a charged particle.

The solutions obtained using the Landau-Lifshitz equation lacks preacceleration, unlike the ones obtained using the Abraham-Lorentz one. However, in case of the rectangular force and to some extent, the exponential force, the Landau-Lifshitz equation seems to fail. The rectangle is arguably an unfair test because $\dot{F}$ blows up at the discontinuities. Perhaps the discontinuity in $\dot{F}$ accounts for the problems with the rectangular and the exponential forces. In some cases the Landau-Lifshitz formula is a good approximation. But it would be an exaggeration to represent the Landau-Lifshitz equation as ``physically correct.''

In spite of these problems, the Landau-Lifshitz equation can be quite useful in practical computations. It differs from the Abraham-Lorentz equation only in second order in $\tau$, and its solutions to realistic problems are practically indistinguishable from those of the Abraham-Lorentz equation since $\tau$ is orders of magnitude smaller than the smallest measurable time interval.

Because the Landau-Lifshitz equation is usually easier to solve, it is often the choice of physicists investigating radiation effects. Its use is well justified for realistic fields that vary slowly on the scale of $\tau$. Even in cases where abrupt changes in the field arise, perhaps as a means of approximating the true field, Landau-Lifshitz solutions should be useful as long as one’s concern is only the particle trajectories and not the radiated fields.

\bibliographystyle{ieeetr}
\bibliography{references}

\end{document}